\documentstyle[psfig]{elsart}
\begin{document}
\begin{frontmatter}
\title{
Analysis of charged particle emission sources and coalescence 
in E/A = 61 MeV $^{36}$Ar + $^{27}$Al, $^{112}$Sn and $^{124}$Sn collisions}
\author{
V.~Avdeichikov$^{a,b,1}$, 
R.~Ghetti$^{a,1}$, 
J.~Helgesson$^{c}$,
}
\author{
B.~Jakobsson$^{a}$,
P.~Golubev$^{a}$, 
N.~Colonna$^{d}$, 
H.W.~Wilschut$^{e}$
}
\address{
 $^a$ Department of Physics, Lund University, Box 118, S-22100 Lund, Sweden\\
 $^b$ Joint Institute for Nuclear Research, 141980 Dubna, Russia \\
 $^c$ School of Technology and Society, Malm\"o University, S-205 06 Malm\"o, Sweden\\
 $^d$ INFN and Dipartimento di Fisica, V.~Amendola 173, I-70126 Bari, Italy\\
 $^e$ Kernfysisch Versneller Instituut, Zernikelaan 25, NL 9747 AA Groningen, 
The Netherlands
}
\begin{abstract}
Single-particle kinetic energy spectra and two-particle 
small angle correlations of protons ($p$), deuterons ($d$) 
and tritons ($t$) have been measured simultaneously in 
61A MeV $^{36}$Ar + $^{27}$Al, $^{112}$Sn and $^{124}$Sn collisions.
Characteristics of the emission sources have been derived from 
a ``source identification plot'' ($\beta_{source}$--$E_{CM}$ plot), 
constructed from the single-particle invariant spectra, and compared to 
the complementary results from two-particle correlation functions. 
Furthermore, the source identification plot has been used to determine 
the conditions when the coalescence mechanism can be applied 
for composite particles. In our data, this is the case only for 
the Ar + Al reaction, 
where $p$, $d$ and $t$ are found to originate from a common source
of emission (from the overlap region between target and projectile).
In this case, the coalescence model parameter, 
$\tilde{p}_0$ -- the radius of the complex particle emission 
source in momentum space, has been analyzed. 
\end{abstract}
\date{\today}
\end{frontmatter}

PACS number(s): 24.10.-i,25.70.Pq, 29.30.Hs \\


{\noindent \small \rm $^1$Corresponding authors. 
Department of Physics, University of Lund. \\ 
Box 118, S-22100 Lund, Sweden. Tel. +46-046-2227647. \\ 
E-mail address: vladimir.avdeichikov@nuclear.lu.se, 
roberta.ghetti@nuclear.lu.se}
\maketitle

\section{Introduction}
\label{sec:intro}

A large body of studies, both experimental 
\cite{Schr92,Pete95,Boug95,Laro95,Rive96,Leco96,Skul96,Dorv99}
and theoretical 
\cite{Sobo94,Sobo97,Eudes,Ditoro1,Ditoro2,Ditoro3,Ditoro4,Ditoro5,Colonna}, 
has demonstrated that intermediate energy heavy ion collisions proceed 
through a complicated reaction mechanism. 
For a wide range of impact parameters, ranging 
from peripheral to near central, the mechanism is dominated by dissipative 
binary collisions, with early dynamical emission followed by statistical 
evaporation. 
In particular, at E/A $\sim$60 MeV, the emission of light particles 
originates from (at least) three sources \cite{Pete95,Lanza98}, 
a quasi-projectile source (QP), a quasi-target source (QT),  
and an intermediate velocity source (IS).
The IS accounts for dynamical emission, described by early 
nucleon-nucleon collisions and by other preequilibrium processes,  
such as emission from a low density neck region 
\cite{Lanza98,Montoya,Toke95,Laro97,Luka97,Pawl98,Laro99,Plag99,Dore00,Mila00,Lefort00,Lanza01,InAl03}. 
Since the particles emitted from these sources overlap in angle, 
energy and emission time, 
it is not possible, on a particle-by-particle basis, to identify which 
source each particle came from. 

A powerful experimental procedure to gain insight into 
the reaction mechanism is two-particle interferometry, 
used to determine the space-time extension of the 
particle emitting sources \cite{Gelbke,Ardouin}, 
and to extract the emission time sequence of neutrons, 
protons and their composites \cite{PRL-01}. 
Our recent interferometry analysis of Ref.\ \cite{PRL-03} on the 
61 MeV/nucleon $^{36}$Ar + $^{27}$Al reaction, has confirmed a dissipative binary 
reaction scenario with an important dynamical emission component from the IS 
created in the overlap zone. 
In particular, an angular dependence of the correlation function 
has emerged. The stronger $pp$ correlation observed at backward angles, 
indicates emission from a shorter lived source.
This has been interpreted as an enhancement of the early dynamical emission 
component seen at backward angles. 
The average emission time of deuterons from this source 
has been found to fall in-between those of neutrons and protons, as expected 
if deuterons are formed mainly by coalescence \cite{Coalescence}.

To complete and complement the interferometry analysis, 
the information contained in the single-particle energy spectra is 
explored in this paper. 
In particular, we investigate the structure of light particle emission 
sources as it follows from  the kinematic features of the single-particle 
({\it p, d, t}) energy spectra. 
A common approach to study single-particle energy spectra, 
is to fit the data to a moving source parameterization, using multiple 
sources to describe preequilibrium and evaporation sources. 
Contribution from three sources is normally assumed: 
a QP, an IS and a QT \cite{Lanza98,Jacak,Wada,Advances,Prindle}. 
While such a phenomenological approach can give a reasonably good
description of double differential particle cross sections, 
the dynamics of the nuclear interaction and the mechanism of particle
production remains hidden in the large number of fitted parameters.
This description is further complicated by the fact that the three-sources 
relative yields vary with impact parameter, so that the parameters utilized 
in the description only represent average values. 
In this paper we investigate an alternative approach to the fitting procedure 
with three moving sources. This approach is feasible when a high-precision 
energy calibration is available, as is the case in the present 
data set \cite{Volly}. 
In this approach, we construct a $\beta_{source}$--$E_{CM}$ plot 
from the invariant energy spectra. Here, $\beta_{source}$ is the 
source velocity in units of $c$, and $E_{CM}$ is the particle 
energy in the source frame.
Inspection of this plot reveals a continuum of source velocities, 
in accordance with the expected trends for a dissipative binary 
collision mechanism of heavy ions. 
For the Ar + Al reaction, the continuum of source velocities 
is found to be common for $p$, $d$ and $t$ charged particles, 
and therefore we can proceed to apply 
a coalescence mechanism, which is proved to be the dominant process of {\it d} 
and {\it t} emission from the overlap region between target and projectile. 
The coalescence model parameter, $\tilde{p}_0$ -- the radius of complex particle 
emission source in momentum space, is analyzed, which provides 
valuable complementary information to the two-particle correlation 
size measurement. 

The paper is organized as follows. Sec.\ \ref{sec:experiment} presents 
the details of the experiment.
Sec.\ \ref{sec:results} presents the source analysis for the 
$^{36}$Ar + $^{27}$Al reaction. 
The angular dependence of $p$, $d$ and $t$ single-particle energy 
spectra is presented in Sec.\ \ref{sec:spectra}; the comparison with 
results from Firestreak model calculations highlights  
some simple physics behind the emission mechanism. 
Sec.\ \ref{sec:beta-tcm} introduces the $\beta_{source}$--$E_{CM}$ plot, 
and discusses the wealth of information it contains on the multiple 
source reaction scenario. In Sec.\ \ref{sec:hbt}, the 
information extracted from the  $\beta_{source}$--$E_{CM}$ plot 
is compared with the results from our previous 
interferometry analysis \cite{PRL-03}. 
Coalescence analysis of the $d$ and $t$ spectra is performed and 
discussed in Sec.\ \ref{sec:coalescence}. 
Sec.\ \ref{sec:tin} extends the $\beta_{source}$--$E_{CM}$ source analysis 
to 61A MeV $^{36}$Ar + $^{112,124}$Sn collisions. 
A summary can be found in Sec.\ \ref{sec:summary}. 
Appendix A describes the details of the Firestreak model calculations.
Appendix B describes the details of the construction and interpretation 
of the $\beta_{source}$--$E_{CM}$ plot.

\section {Experimental details}
\label{sec:experiment}

The experiment was performed at the AGOR cyclotron of KVI (Groningen).  
The E/A = 61 MeV $^{36}$Ar$^{14+}$ pulsed beam ($\sim$ 0.5 $\div$ 1.0 nA) 
impinged on targets of $^{27}$Al, $^{112}$Sn and $^{124}$Sn 
(1.8 mg/cm$^2$ thick). The beam intensity was monitored by a Faraday cup.
The experiment measured simultaneously single-particle 
energy spectra and small angle two-particle correlations, in
coincidence with forward emitted fragments. 
The requirement of at least one fragment detected in the 
forward direction biases our collected data 
towards semi-peripheral collisions \cite{NPA}.

The particles were detected using three complementary multidetector 
systems: the EMRIC charged particle detector array \cite{Volly,Merchez}, 
the EDEN neutron detector array \cite{EDEN,NIMTOF}, 
and the KVI Forward Wall (FW) phoswich detector \cite{WALL}.   
The experimental data were collected in two modes. In the first
mode, both EMRIC and EDEN registered inclusive particle spectra
in coincidence  with the FW. In the second mode, 
coincidence events between EMRIC, EDEN and EMRIC*EDEN elements 
were collected in coincidence with the FW. 

The EMRIC charged particle detector array consisted of 16 CsI(Tl)
elements, each one subtending a solid angle of 3.6 or 5.1 msr. 
The detector modules were arranged in two groups, consisting of 
single elements and triple elements. 
The complete geometry, optimized for the interferometry requirements, 
is presented in Ref.\ \cite{NIMTOF}.
All detectors were operated in air, outside a scattering
chamber of (3 mm thick) stainless steel and 28 cm in radius.
100 $\mu$m thick Capton foil windows were installed in front 
of the EMRIC elements. One extra element,
equipped with a $^{241}$Am $\alpha$-source and placed at 30$^{\rm o}$
was used to monitor the shift in the $\alpha$-peak position caused
by the variation of beam intensity and, consequently, the
$\gamma$-background, detected by all EMRIC elements.
Temperature monitoring of the CsI(Tl) elements 
was also applied \cite{Volly}. 
The detectors and measuring technique are described in details
in Ref.\ \cite{Volly}. The $^{1,2,3}$H isotopes were
identified via the pulse-shape analysis method. 
An accuracy $<$ 1\% was achieved in the energy
calibration for each of the 100 mm long crystals by using the
$\Delta$E(Si)--E(CsI(Tl))/PMT method. The validity of the
calibration was checked by comparing the particle spectra
measured by detectors located at the same polar angle. 
Isotopic resolution was obtained for $Z$=1
starting from energy threshold $\approx$ 8, 11 and 14
MeV, for {\it p, d} and {\it t}, respectively.

Two important corrections, both energy dependent, were introduced in
the measured particle spectra. These corrections are quite common
for experiments with long scintillators.
First, the correction on the loss of charged
particles due to their inelastic interaction with the
scintillator material. Experimental data and calculations 
\cite{VAv437} may be approximated with a good precision by
the equation,

\begin{equation}
k_{inel}=1 + b(Z,A)E[1 - exp(-E/100)],
\label{eq:1}
\end{equation}
where $E$ is the particle kinetic energy in MeV, $b(Z,A)$ is a fitting
parameter. 

Second, the correction on the change in real solid angle, $\Omega_{real}$, of
the scintillator along its length,

\begin{equation}
k_\Omega = \Omega_{input}/\Omega_{real} = (1 + \frac{c(Z,A)}{L}E^{1.68})^2.
\label{eq:2}
\end{equation}
Here $\Omega_{input}$ is the solid angle defined by the front face
active area of the scintillator, $L$(cm) is the distance from the target
to the scintillator front face, and $c(Z,A)$ is a fitting coefficient, 
extracted from the energy-range relation for particle
($Z,A$) in CsI crystal \cite{TRIM95}\footnote{The coefficients $b(Z,A)$ and $c(Z,A)$ 
are equal to 0.0014, 0.00090, 0.00066 and 
0.00140, 0.00155, 0.00115 for {\it
p, d, t}, respectively.}. 
The measured spectra have been corrected by
the factor $k = k_{inel} \cdot k_{\Omega}$, 
i.e.\ $N(actual,E) = k \cdot N(measured,E)$.
 As an example, the total correction 
factor for protons is calculated to 1.76 for $E$=200 MeV.

\section{Source analysis for E/A = 61 MeV $^{36}$Ar + $^{27}$Al collisions}
\label{sec:results}

\subsection{Single-particle energy spectra}
\label{sec:spectra}

Proton, deuteron and triton energy spectra, produced in 61 MeV/nucleon
$^{36}$Ar + $^{27}$Al collisions, were measured in the energy intervals 
$\sim$ 8(11)(14)--250 MeV for 8 angles in the polar angle region  
30$^{\rm o}$--114$^{\rm o}$. Fig.\ \ref{spectra} presents 
proton (upper panel), deuteron (middle panel) and triton 
(bottom panel) data for some laboratory angles, as listed in each panel.

We have compared the proton data (Fig.\ \ref{spectra}, upper panel) 
to the results of a Firestreak model calculation (histograms) 
and normalized to the data at 78$^{\rm o}$. 
The basic assumptions of the Firestreak model are reviewed 
in Appendix A. This model uses a quite drastic approximation 
of the geometry and treats the overlap region of the target 
and projectile as created in a totally inelastic process.
Yet, a reasonably good description of the angular and energy
distributions of the protons is achieved.
It should be stressed that only one parameter -
the maximum source temperature, $T_{source}$=13.2 MeV 
is introduced. 
The quite good agreement with the experimental data 
shows that the nature of the binary dissipative
collisions is closely related to the participant-spectator 
picture of the interaction (see also Sec.\ \ref{sec:beta-tcm}). 

\subsection{Source identification, $\beta_{source}$--$E_{CM}$ plot}
\label{sec:beta-tcm}

The simplest way to describe the
particle spectra is to adopt the participant-spectator 
description and assume emission from three 
thermal moving sources. 
For each emitting source, the velocity,
$\beta_{source}$, source temperature, $T_{source}$, Coulomb
barrier, $V_{c}$, and one normalization constant are treated as 
free parameters. At least 12 parameters must be determined by a
simultaneous fit to the particle spectra at the measured angles. To
overcome the problem  of coupling between the parameters, some of
them must be kept fixed. Such a parametrization gives a reasonably 
good description of the particle spectra and angular
distributions \cite{Lanza98,Jacak,Wada,Advances,Prindle}.  

As an alternative way we construct a ``source identification plot''
from invariant cross sections. 
We obtain the source velocity, $\beta_{source}$, and 
the kinetic energy of the particles in the source frame, 
$E_{CM}$, from invariant cross sections, 
using different lab energies and different angular combinations.
In this way, we construct a $\beta_{source}$--$E_{CM}$ plot 
as if only one source was responsible for emission of particles.
When several sources are present, the extracted value 
of $\beta_{source}$ represents an 
effective average source. However, and this is an 
important point, the effective source will have 
different characteristics, depending on the angles 
of the chosen detector pairs. By careful analysis of 
the source identification plot, the properties of the 
contributing sources may be revealed to a certain extent. 
In Appendix B we discuss the derivation and properties of such
$\beta_{source}$--$E_{CM}$ plots further. 
Fig.\ \ref{beta-tcm} presents the results
from the $\beta_{source}$--$E_{CM}$ method with 
data from $^{36}$Ar + $^{27}$Al collisions. 
Protons (full symbols), deuterons (big open symbols) 
and tritons (small open symbols) have been studied. 
The results are shown for some representative pairs 
of angles (selected for their highest statistics, 
taken with the triple EMRIC detectors).
The value of $E_{CM}$ is plotted in MeV/nucleon. 
The solid lines are drawn to guide the eye. 
Values of the source velocity corresponding to total fusion 
($\beta_{source}$ = 0.205) and to the projectile velocity 
($\beta_{source}$ = 0.361) are marked by the arrows on the $y$-axis. 

This plot contains various types of information, 
which we now proceed to discuss. 

1) A main feature is that, in the intermediate region of $E_{CM}$, 
there is a clear spreading of values of $\beta_{source}$. 
This is due to the contribution from different sources (QT, QP and IS) 
and to the broad impact parameter range of our apparatus, 
leading to a continuum of source velocities, 
covering the region 0.07 $ < \beta_{source} < $0.22 (common 
to all charged particles). This is in accordance with the expected
trends for a dissipative binary collision of heavy ions. 
The observed continuum of emission sources is quite naturally
reproduced in the nuclear Firestreak model description discussed 
in Sec.\ \ref{sec:spectra} and in Appendix A (Fig.\ \ref{Firestreak}). 

2) One can identify a limiting region at high $E_{CM}$, 
where the source velocity converges 
towards the value $\beta_{source}\sim$~0.21.
This indicates that the particles emitted at different angles, 
with the highest kinetic energy, originate from a source (IS) 
with a well defined velocity.

3) Another limiting region can be identified 
at low $E_{CM}$ and backward angles. Here the different 
lines tend to converge towards the QT velocity. 
These low kinetic energy particles may be identified as 
emitted by the QT source (compare with Fig.\ \ref{tano}).

4) The ``bump'' observed for low $E_{CM}$ and forward angles 
(the 30$^{\rm o}$/54$^{\rm o}$ pair), represents a third limiting region. 
This ``bump'' indicates that particles with 
a quite low kinetic energy are emitted by a source with a high velocity. 
These particles may be identified as emitted by 
the QP source (compare with Fig.\ \ref{tano}).
This region is not completely explored by our experimental 
apparatus, due to the limited coverage at forward angles.
Notice that the QP ``bump'' disappears if the source 
identification plot is constructed for particles 
detected in coincidence with 4 or more fragments 
in the FW, a condition that biases the collected events 
towards less peripheral collisions. 

5) The {\it p, d, t} particles are emitted by common sources 
(i.e.\ the corresponding points lie on the same line in the plot). 
This is a key point that justifies the application of the coalescence 
mechanism of {\it d} and {\it t} production 
(see Sec.\ \ref{sec:coalescence}) and, maybe, a unique lifetime 
for their emission. 

\section{Comparison with interferometry results 
for E/A = 61 MeV $^{36}$Ar + $^{27}$Al collisions}
\label{sec:hbt}

The source characterization from the $\beta_{source}-E_{CM}$ 
study, can be compared with the interferometry 
analysis carried out in Ref.\ \cite{PRL-03}. 
There it was found that the correlation strength 
of $pp$ pairs measured in the ``forward'' angular 
region (54$^{\rm o}$/30$^{\rm o}$) is smaller 
than the correlation strength measured backwards.
This indicates a larger space-time dimension of the effective 
source that emits protons in the forward direction. 
This is in agreement with the $\beta_{source}-E_{CM}$ result 
that detectors at ``forward'' angles, 54$^{\rm o}$/30$^{\rm o}$, 
register particles originating 
from a superposition of two sources, 
the IS and the QP source \cite{Eudes}. 
The QP decay reveals itself as the ``bump'' 
at $E_{CM} \sim$ 25 MeV/nucleon.

In contrast, an effective source of smaller space-time size was 
deduced from the stronger $pp$ correlation function measured at 
backward angles \cite{PRL-03}. 
This result seems to be in contrast with the source identification 
plot in Fig.\ \ref{beta-tcm}, where at backward angles 
(78$^{\rm o}$/102$^{\rm o}$) a QT can be identified 
at low $E_{CM}$ energies. However, one should notice that in the 
correlation function, the yield from the different sources 
(an information not seen in the $\beta_{source}-E_{CM}$ plot)
plays an important role. Since the measured yield from the 
QT source is low, the dominating contribution to the correlation 
function actually comes from the region of high $E_{CM}$ energies 
in the source identification plot, dominated by the IS.

Finally, also the results of the $pp$ correlation functions 
with a high total momentum cut ($P_{tot}> $ 450 MeV/c) are 
in agreement with the information derived from the 
$\beta_{source}$--$E_{CM}$ plot. 
In Ref.\ \cite{PRL-03} it was found that the 
high total momentum gated $pp$ correlation functions 
are enhanced and similar in size at backward and forward angles, 
indicating emission from an IS of small space-time dimension. 
The $pp$ correlation function measured forward is more substantially 
enhanced. This is because the momentum gate at forward angles 
effectively corresponds to suppressing the contribution from the QP source, 
leaving the IS at high $E_{CM}$ (high $P_{tot}$) as the dominating source.
At backward angles, on the other hand, the effect of the high 
total momentum gate is only a small enhancement, since the contribution from the 
QT source was already small (see discussion above). 

\section{Coalescence analysis for E/A = 61 MeV $^{36}$Ar + $^{27}$Al collisions}
\label{sec:coalescence}

The common sources for {\it d} and {\it t} emission, Fig.\ \ref{beta-tcm}, 
suggest their common origin, f.e., through the coalescence mechanism. 
Calescence models are reviewed in Ref.\ \cite{Coalescence}.

An empirical nucleon coalescence model assumes a simultaneous
emission of protons and neutrons (whose relative momentum is less
than a certain value $\tilde{p}_{0}$) by the source to form a
complex particle ($Z,A$). 
Indeed, our study of the particle emission time sequence 
via two-particle correlation functions \cite{PRL-03} 
is consistent with this assumption 
(as discussed in  Sec.\ \ref{sec:intro}).
The invariant momentum spectrum (or 
momentum space density) of the composite particle can be expressed 
through the neutron and proton invariant spectra,
\begin{equation}
E_{A}d^{3}
\sigma_{A}/d^{3}p_{A}=B_{A}(E_{p}d^{3}\sigma_{p}/d^{3}p_{p})^{Z}
\cdot(E_{n}d^{3}\sigma_{n}/d^{3}p_{n})^{N}.
\label{eq:4}
\end{equation}
Here $p_{A}$=Ap$_{p,n}$. $E_{p}$, $E_{n}$ and $E_{A}$ are the
proton, neutron and composite particle total energies\footnote{ 
Because of experimental restrictions, the spectra of protons and
neutrons were registered by our apparatus at slightly different 
angles. Therefore, in the coalescence analysis we assume that 
they are equivalent for a given angle, but scaled by a factor
 $R_{n,p}=(N_{t} + N_{p})/(Z_{t} + Z_{p})$, 
corresponding to the ratio of neutron to proton numbers in 
the projectile and target.}. 
The ``coalescence factor'' is given in the classical approximation as:
\begin{equation}
B_{A}=A \cdot \frac{2S_{A}+1}{2^{A}} \cdot
\frac{1}{Z!N!}(\frac{4\pi
\tilde{p}_{0}^{3}}{3m\sigma_{0}})^{A-1}\cdot R_{n/p}^{N}.
\label{eq:5}
\end{equation}
Here $m$ is the nucleon mass, $S_{A}$ is the spin of the particle,
$\sigma_{0}$ is the ``hot zone'' formation cross section, and
$\tilde{p}_{0}$ is the ``coalescence radius'' in momentum space.
The value of $\sigma_{0}$ cannot be defined unambiguously.
Usually, it is taken as the geometrical reaction cross section.
The choice of $\sigma_{0}$ influences the extracted
$\tilde{p}_{0}$ value.

If chemical and thermal equilibrium are achieved in the
``hot zone'' volume \cite{Mekjan}, $\tilde{p}_{0}$ can be related
to the volume $V_{0}$ within which the nucleons coalesce to form
a composite particle,
\begin{equation}
 V_{0} \sim \frac{Z!N!}{\tilde{p}_{0}^{3}}.
\label{eq:6}
\end{equation}
For the dissipative binary type of collisions studied here,  
charged particles are emitted mostly in preequilibrium processes 
(with emission time $<$ 50 fm/c \cite{Eudes,PRL-03}). 
Thus we take $\tilde{p}_{0}$ 
only as a parameter of the coalescence model.

Coalescence model calculations of {\it d} and {\it t} spectra are
presented in Fig.\ \ref{spectra} for some
representative angles. 
The histograms drawn for {\it d} (middle panel) and $t$ 
(bottom panel) represent the calculation in 
coalescence model with $\tilde{p}_{0}$= 100 and 135 MeV/c for {\it d} 
and {\it t}, respectively. Agreement between the model calculation 
and the data is very good, with the exception of the 30$^{\rm o}$ angle, 
where the low-energy part of the spectrum comes from the QP decay. 
(It should be noted here that the coalescence analysis must be restricted 
to the kinematical region for which particle emission from the
QP and QT sources is negligible).

Quantum mechanical treatment of the coalescence mechanism \cite{Sato} 
assumes that the momentum distributions of the emitted particles
can be described by density matrices. The coalescence volume is
related to the internal wave function of the composite particle
and of the spatial (Gaussian) distribution of nucleons in the
source. Notice that the model does not impose thermal or chemical 
equilibrium. The coalescence radius is expressed as:
\begin{equation}
\tilde{p}_{0}(A) \approx
const(N,Z,A,S)[\frac{1}{R^{2}_A}+\frac{1}{R^{2}_0}]^{1/2},
\label{eq:8}
\end{equation}
where $R_{A}$ and $R_{0}$ are the radii of the composite particle 
and of the emitting source, respectively.  
The source radius extracted by applying the density matrix 
coalescence model is $\sim$ 3.5 fm, both for deuterons and tritons. 

\section{Source analysis for E/A = 61 MeV $^{36}$Ar + $^{112,124}$Sn collisions}
\label{sec:tin}

Fig.\ \ref{beta-tcm-sn} presents the source identification plot 
$\beta_{source}$--$E_{CM}$ constructed from 
the invariant spectra of protons, 
deuterons and tritons emitted in 61A MeV $^{36}$Ar + $^{112}$Sn (upper panel) 
and $^{36}$Ar + $^{124}$Sn (lower panel) collisions.

Inspection of the plots reveals that:

1) A source velocity continuosly varying with $E_{CM}$ 
is again extracted (as for the Ar + Al system), 
in the region of 0.03 $ < \beta_{source} < $0.2. 

2) The results for the two Sn targets are very similar, 
which indicates very little isospin dependence in the source 
velocities. A somewhat larger contribution of the target 
residue is noticeable at backward angles in the Ar + $^{124}$Sn system.

3) The  Ar + Sn results do not show a QP limiting region. As compared to 
the Ar + Al plot, the ``bump'' observed in the $\beta_{source}$--$E_{CM}$ 
dependence for low $E_{CM}$, forward angles, is absent in  Ar + Sn. 
This indicates that our experimental apparatus misses the 
QP component in the direct kinematics Ar + Sn reactions, due to 
lack of forward angle coverage.

4) The sources are common for $p$, $d$ and $t$ at high energies 
($E_{CM}$~$>$30 MeV/nucleon). In the low $E_{CM}$ range, instead, 
protons deviate from deuterons and tritons. Inspection of this deviation 
reveals that the QT component is more sizable for proton emission. 
This implies that a coalescence model analysis of the deuteron and triton 
production is not feasible in Ar + Sn, since, as it was pointed out in 
Sec.\ \ref{sec:coalescence}, the coalescence analysis must be restricted 
to the kinematical region for which particle emission from the
QP and QT sources is negligible.

\section{Summary}
\label{sec:summary}
We have proposed a novel and simple method to analyze
single-particle energy spectra. 
This method allows to characterize the multiple sources of 
particle emission that are present in dissipative binary, 
intermediate energy heavy ion reactions.
The method consists in constructing a source identification 
plot $\beta_{source}$-$E_{CM}$. 
If the accuracy of the energy calibration is
sufficiently high, this plot can be generated using the
invariant cross section measured
at different laboratory angles.

Inspection of this plot suggests that 61 MeV/nucleon Ar + Al, Sn
collisions are characterized by sources with continuously 
varying velocities, that tend to converge towards a 
well defined source velocity for particles emitted with high $E_{CM}$. 
Thus, this analysis supports 
a dissipative binary reaction scenario, with
an important contribution from an intermediate velocity
source. 

The results from the Ar + Al collisions have been compared
with the two-particle correlation function analysis performed
for coincidence data from the same experiment \cite{PRL-03}.
A connection between the apparent source size (containing 
space-time contributions from different sources) extracted from
the correlation functions and the results from the
$\beta_{source}$-$E_{CM}$ plot has been established.

The $\beta_{source}$-$E_{CM}$ plot also allows to compare
the origin of emission of particles of different types, and
to reliably establish whether a common origin, necessary
condition for the applicability of a coalescence mechanism,
is met. In our data, we have found that this is the case for 
the Ar + Al collisions, but not for Ar + Sn. 
Application of the coalescence model to the Ar + Al data, 
has allowed to extract a coalescence radius, common for 
deuterons and tritons.
\\

{\bf ACKNOWLEDGEMENTS}\\

The authors wish to thank the European Commission support in
the framework of the Transnational Access Program under contract
HPRI-CT-1999-00109. Financial support from the Swedish 
Research Council under contracts F 620-149-2001 (RG) and 
621-2001-1782 (BJ) is also acknowledged. 

\vspace{1.cm}
{\bf APPENDIX A : Firestreak Model}\\

The Firestreak model introduces a possibility to describe 
the fluctuations in souce size and excitation energy 
of the strongly interacting fireball. 
In our approach we follow essentially the prescription given 
in Ref.\ \cite{VolAvd}. A nuclear density distribution is 
incorporated for both colliding nuclei. 
The collision is treated as a totally
inelastic process in the overlap region of the target and
projectile. The interaction proceeds via collinear streaks of
nuclear matter of both nuclei. Each of the streaks is
characterized by the value of ``projectile fraction'',

\begin{equation}
\eta = n_{p}/(n_{p} + n_{t}),
\label{eq:9}
\end{equation}
where $n_{p}(n_{t})$ is the number of contributing nucleons from
the projectile (target). The local value of source velocity,

\begin{equation}
\beta (\eta) = \eta \cdot \beta_{beam},
\label{eq:10}
\end{equation}
and the internal energy per nucleon,

\begin{equation}
t(\eta) = \eta(1 - \eta)t_{beam},
\label{eq:11}
\end{equation}
are both expressed through the value of $\eta$. Here
$\beta_{beam}$ is the velocity of the projectile, and t$_{beam}$
is the kinetic energy per nucleon of the projectile. The
geometrical aspects of the interaction (in units of cross
section) are contained in the ``yield function'', Y($\eta$).

\begin{figure}
\vspace{3.5cm}
\centerline{\psfig{file=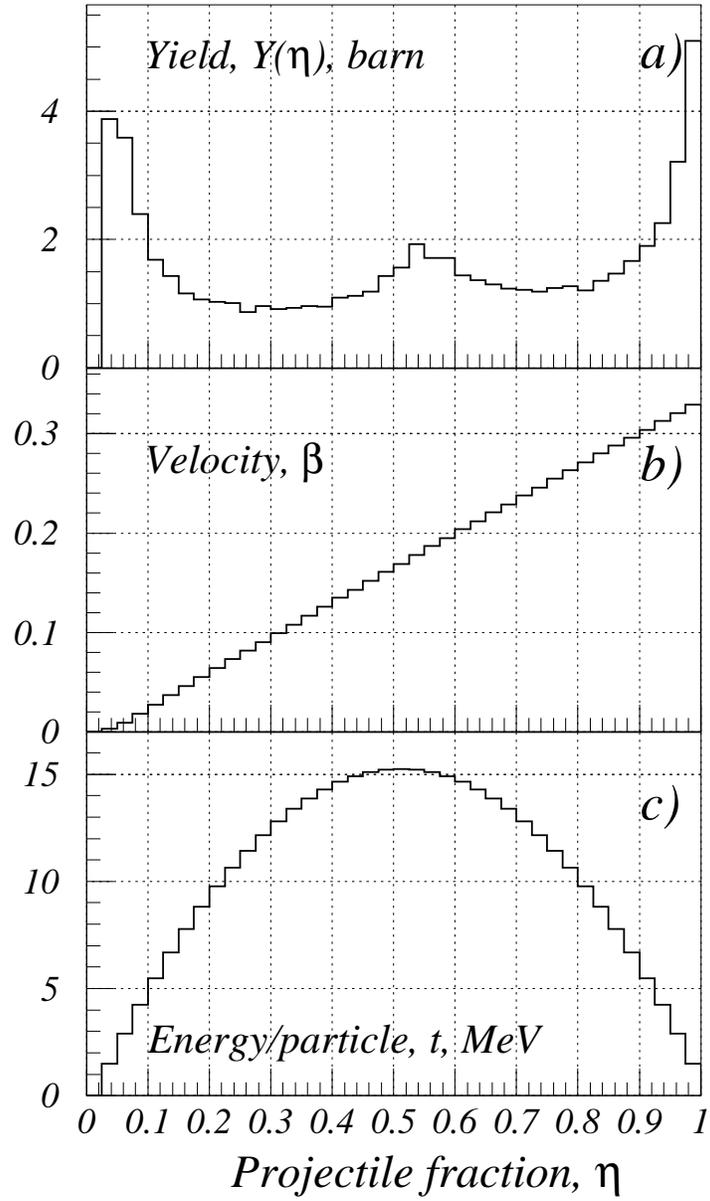,height=16cm,angle=0}}
\vspace{-2cm}
\caption{\small
  Summed over impact parameter (with $b_{max}$= 12.7 fm) the yield function, 
  $Y(\eta$), the velocity, $\beta$, and internal energy per particle, {\it t}, 
  are plotted against the projectile fraction $\eta$
  for 61 MeV/nucleon $^{36}$Ar + $^{27}$Al collisions.
}
\label{Firestreak}
\end{figure}

Fig.\ \ref{Firestreak} presents the results of our numerical 
calculation for $^{36}$Ar + $^{27}$Al collisions at 61A MeV. 
We used a three-parameter Fermi-type nuclear density distribution 
$\rho (r)$ for both colliding nuclei. The parameters are taken as 
compiled in Ref.\ \cite{Bar}. 
A parabolic interpolation was used to get parameters for
the nuclei of interest in this work.  The bins of Y($\eta$) near
$\eta \approx$ 0 correspond to contributions from the target
spectator, with quite low internal (excitation) energy and
source velocity. The bins of Y($\eta$) near $\eta \approx$ 1
correspond to contributions from the projectile spectator, with
quite low internal energy but with high source velocity. 

\begin{figure}
\centerline{\psfig{file=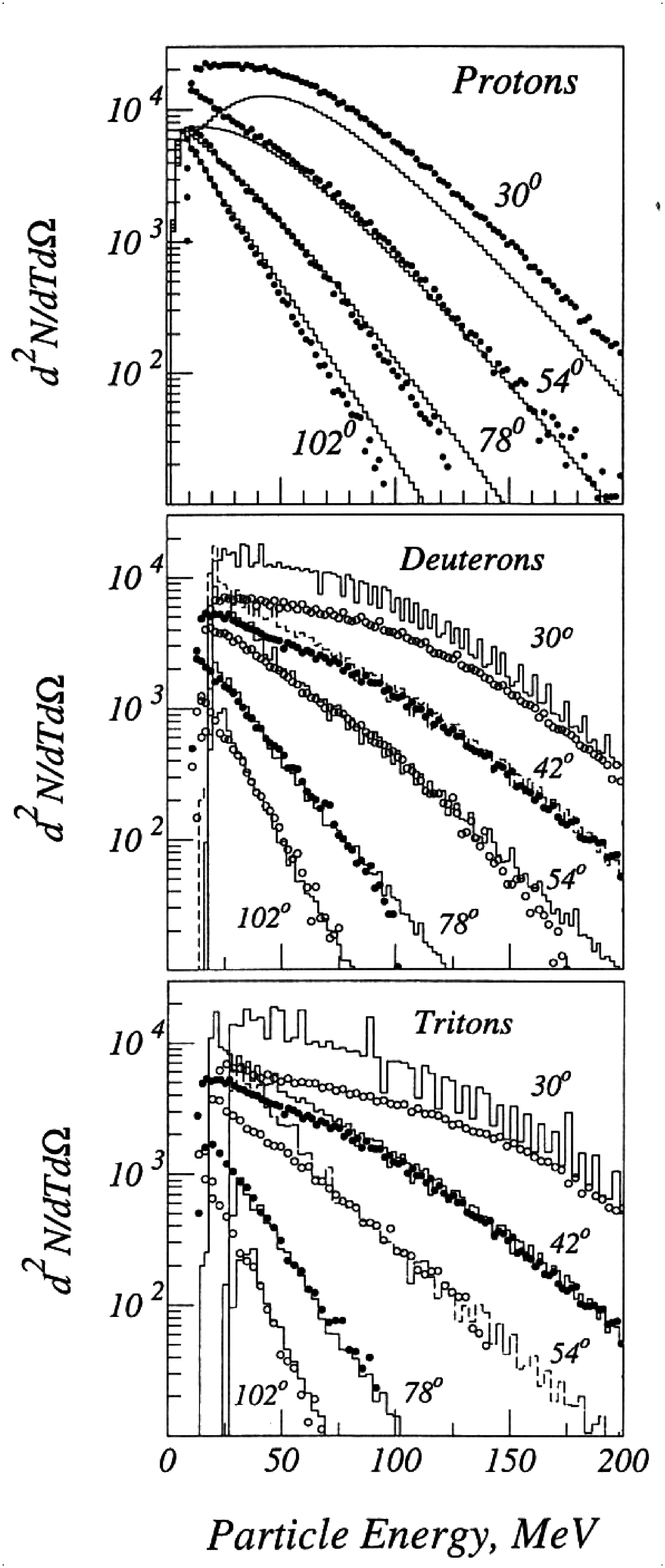,height=20cm,angle=0}}
\caption{\small
  Proton, deuteron and triton Lab energy spectra produced in 61 MeV/nucleon
  $^{36}$Ar + $^{27}$Al collisions, for some angles as listed
  in each panel. The data are the open and filled circles.
  The upper panel represents protons (points) and Firestreak 
  model calculations (histograms), normalized to data at 78$^{\rm o}$.
  Coalescence model calculations (Sec.\ \protect{\ref{sec:coalescence}}) 
  for deuterons (middle panel) and tritons (bottom panel) are plotted as histograms. 
  The $d^{2}N/d\Omega dE$ scale is presented in $\mu b/(sr \cdot MeV)$.
}
\label{spectra}
\end{figure}

The Lab momentum space density for particle of type {\it j} is
given by the expression,
\begin{equation}
F_j({\bf p})=\sum_{i=1}^{n}Y(\eta_i)J_{{\bf p^{'}} \rightarrow
{\bf p}}[\beta(\eta_i)] \cdot f_j[{\bf p^{'}},t(\eta_i)],
\label{eq:12}
\end{equation}
where J$_{{\bf p^{'}} \rightarrow {\bf p}}$ is the Jakobian
transformation from the CM to Lab frame, and $f_j[{\bf
p^{'}},t(\eta_i)]$ is the CM frame momentum distribution for
protons. Since we  apply here  only a phenomenological description
of the proton spectra, the choice of $f_{proton}[{\bf
p^{'}},t(\eta_i)]$ is quite arbitrary. Furthermore, we do need the 
assumption that particles in the local source have come to
thermal equilibrium in their CM frame. So we  accept a classical 
Maxwell momentum distribution in the CM frame of the emitting system,
\begin{equation}
f_{proton}({\bf p^{'}},t)\sim
(1/t^{2})exp(-\frac{p^{'2}/2m}{2t/3}),
\label{eq:13}
\end{equation}
with a surface proton emission. The Coulomb barrier is taken in
the source frame, $V_{c}$=2.0 MeV. Results from the calculations are
presented in Fig.\ \ref{spectra} (protons). To get consistency with data,  
the ``effective'' temperature, $2t(\eta_{i})/3$, is increased by
factor $\alpha$=1.32.
If one postulates, as in the hydrodynamical approach \cite{Beck},
that the nuclear fluid is viscous, then friction arises over
the region of overlap of the target-projectile. Work done
against this friction force is transformend into a thermal energy, 
which essentially
increases the participant and spectators excitation energy
\cite{VolAvd} and this motivates the $\alpha$-factor.\\

{\bf APPENDIX B: Source velocity
reconstruction by the $\beta_{source}$--$E_{CM}$ method}\\

To facilitate the reconstruction of the sources we use
different detector combinations to extract source
velocity, $\beta_{source}$, and particle energy 
in the source frame, $E_{CM}$, from the invariant spectrum.
If there is only one unique emission source from a reaction, 
the invariant cross sections at different angles 
can be used to determine the source velocity. 
The Lab energies $E_1$ and $E_2$ of 
particles emitted at different angles with equal  
invariant cross sections should thus coincide in energy ($E_{CM}$) 
after a transformation to the source system, moving with 
$\beta_{source}$ in the laboratory. 
This allows to extract the value of $\beta_{source}$ 
and $E_{CM}$ from the shift in particle kinetic energy,
$E_{1}-E{2}$, for each pair of Lab angles 
$\theta_{1}$ and $\theta_{2}$ using two equations,
\begin{equation}
  E_{CM} = E_{1} -2(E_{1}\cdot E_{source})^{1/2} \cdot
           cos\theta_{1} + E_{source}.
\label{eq:2-12}
\end{equation}
\begin{equation}
  E_{CM} = E_{2} -2(E_{2}\cdot E_{source})^{1/2} \cdot
           cos\theta_{2} + E_{source}.
\label{eq:2-13}
\end{equation}
Here $E_{source}$={\it m}$\beta_{source}^{2}$/2 is the Lab kinetic
energy of particle {\it m}, having the velocity $\beta_{source}$.
In fact, the described procedure means extracting $E_{CM}$ and
$\beta_{source}$ by restoring the hypothetical invariant cross
section in the source frame by using Eqs.\ (\ref{eq:2-12}, \ref{eq:2-13}).
It should be noticed that for this method to yield reliable 
results, the energy calibration must be known with a high 
accuracy. For example, an accuracy $\pm 3 \%$ in the energy 
$(E_1,E_2)$ yields approximately $\pm 10 \%$ error in the 
extracted $\beta_{source}$.

By combining detectors at different laboratory angles
and using particles emitted with different energies
a number of different combinations of 
$\beta_{source}$--$E_{CM}$ can be obtained.
The method described above assumes only one
source present.
When several sources are present, the extracted values 
of  $\beta_{source}$ and  $E_{CM}$ represent an 
effective or average source. 

\begin{figure}
\centerline{\psfig{file=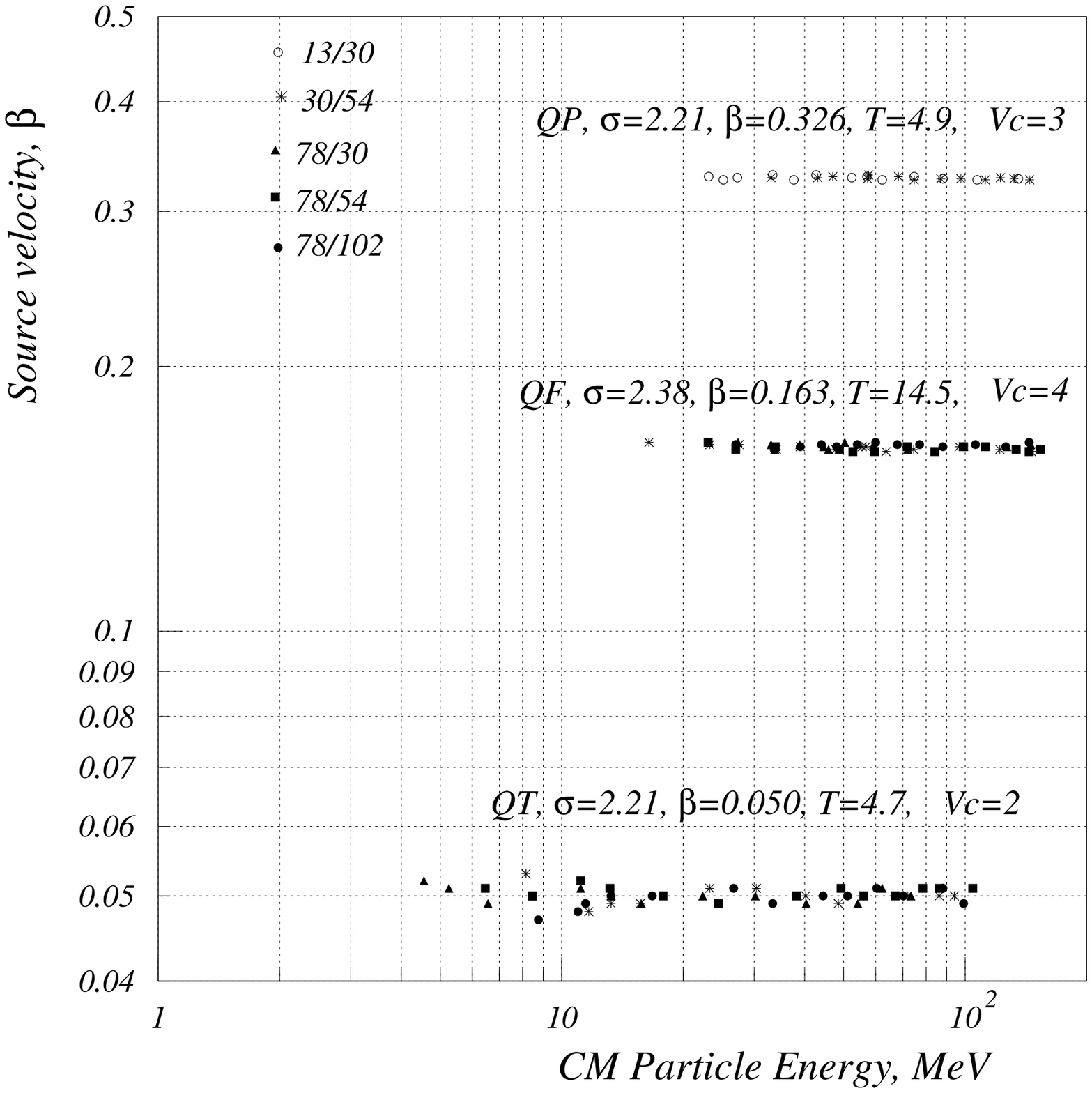,height=7cm,angle=0}}
\centerline{\psfig{file=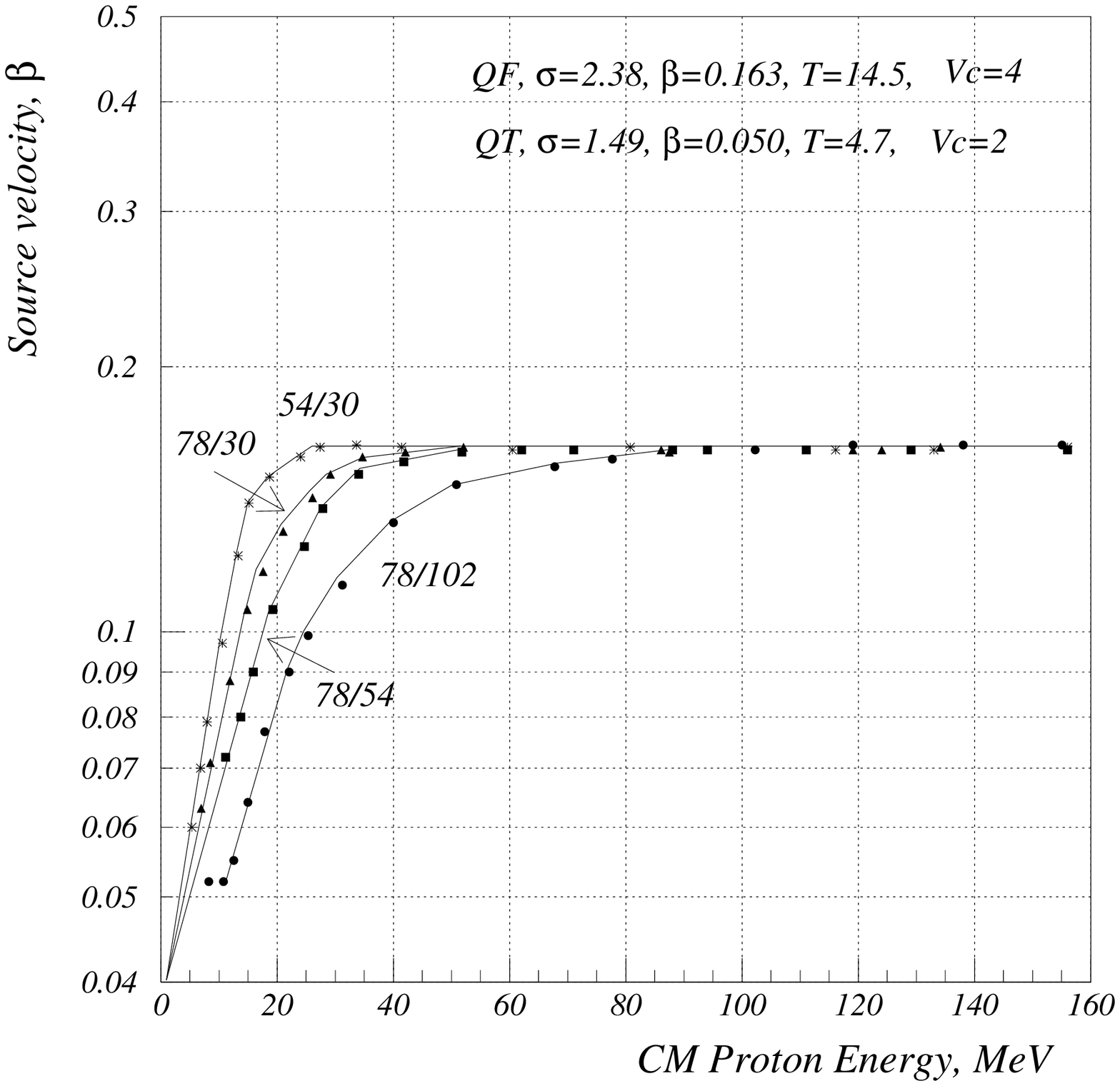,height=7cm,angle=0}}
\centerline{\psfig{file=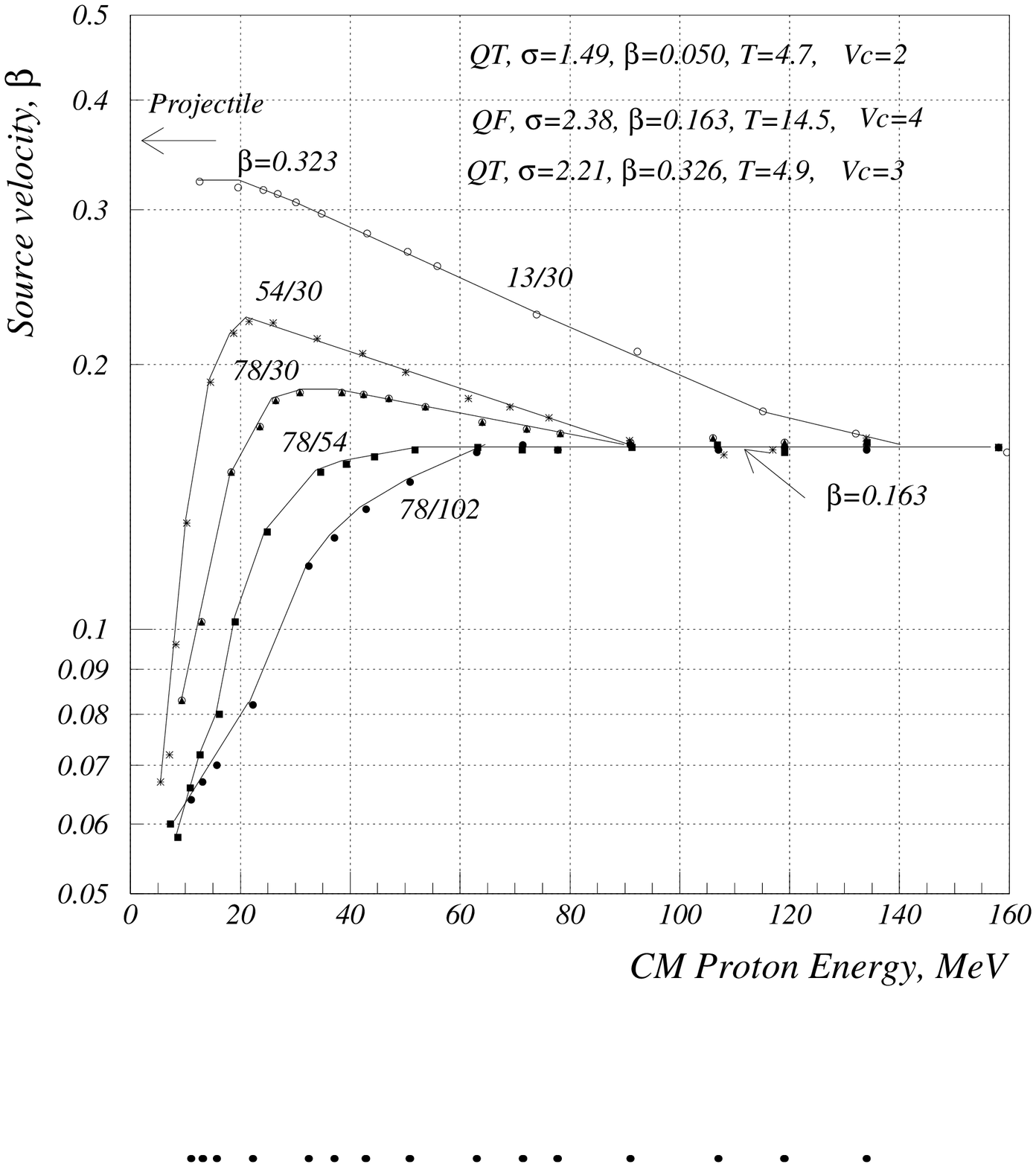,height=7cm,angle=0}}
\vspace{-1.cm}
\caption{\small
  The ``effective'' source velocity $\beta_{source}$ versus the  
  proton kinetic energy in the source frame, 
  as reconstructed from the energy spectra 
  assuming particle emission from single (upper panel), 
  two (middle panel) and three (bottom panel) moving sources. 
  The parameters of the Maxwell-Boltzmann
  distribution in the source frame are listed in the figure 
  (QP is quasi-projectile source, QT is quasi-target source and 
   QF is the intermediate velocity source, in the notation of 
    Ref.\ {\protect\cite{Lanza98}}). 
}
\label{tano}
\end{figure}

To correctly interprete the $\beta_{source}$ and  $E_{CM}$
obtained for our experimental data, we construct
a $\beta_{source}$ and  $E_{CM}$ from a known input.
For this purpose we use the three source fit by
G.\  Lanzan\`{o} {\it et al.} \cite{Lanza98},
which has been obtained from the
double differential cross sections for protons emitted in 
E/A = 60 MeV $^{36}$Ar + $^{27}$Al collisions
measured in the angular range 12.5$^{\rm o}$--112.5$^{\rm o}$. 
A standard way to describe the proton spectra and their
angular dependence has been applied, by assuming particle
emission from three thermal moving sources. In the source
reference frame, particle evaporation is described as 
a volume emission of Maxwell-Boltzmann type. 
For each source, the values of
$\beta_{source}$, T$_{source}$ (MeV), V$_{c}$ (MeV) and a normalization
constant, $\sigma$ (in barn), are treated as free parameters. 
The corresponding fit parameters for 3 types of moving sources, 
QT, IS (denoted QF in \cite{Lanza98} 
and in Fig.\ \ref{tano}) and QP are 
determined by a simultaneous $\chi^{2}$ fit to the proton spectra at 
the 13 measured angles. The fit parameters, as deduced by the 
authors \cite{Lanza98}, are listed in Fig.\ \ref{tano}.

We used these parameters to calculate the invariant proton spectra. 
We then applied the procedure described to reconstruct
$\beta_{source}$. The results are presented in Fig.\ \ref{tano}. 
The upper panel presents the results when particle emission from 
one single source is assumed (QP, IS and QT are illustrated in the
same plot). The figure shows that the reconstructed $\beta_{source}$ 
is in general equal to the input values, and the small deviations of some
points illustrates the accuracay of the method. Another important
obeservation is the range of energies populated by the different
sources. Because of kinematical reasons the fast moving QP cannot
populate low energies\footnote{Low and high energy cutoffs 
in the $\beta_{source}$ values are defined by the possibility 
to extract simultaneously $E_1$, $E_2$ from the invariant 
spectra for the chosen pair of angles.}. 
This is important to have in mind to
correctly interprete the ``effective averaging'' when several sources
are present as in the middle and lower panels of Fig.\ \ref{tano}. 

\begin{figure}
\vspace{1.cm}
\centerline{\psfig{file=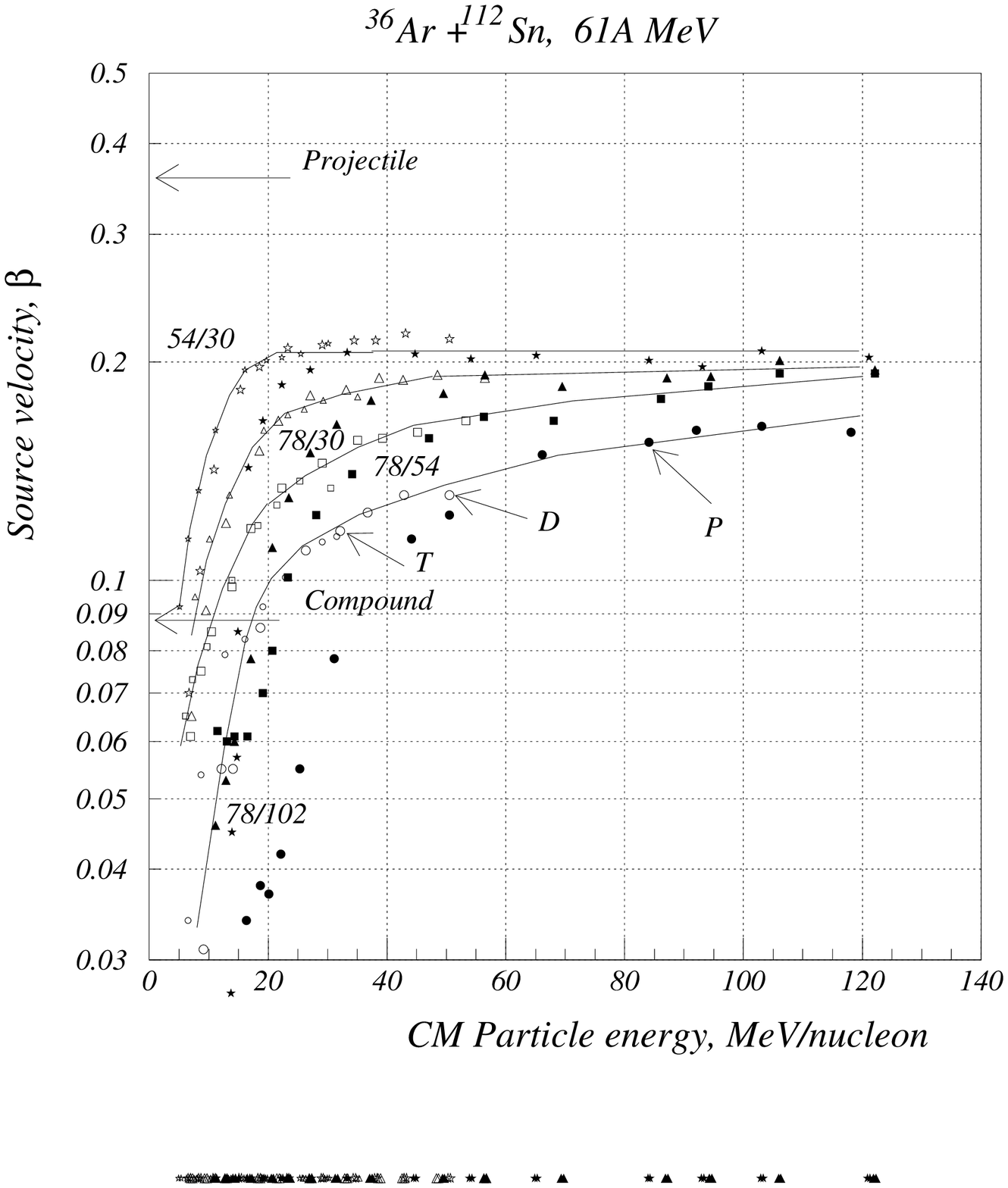,height=10cm,angle=0}}
\vspace{.8cm}
\centerline{\psfig{file=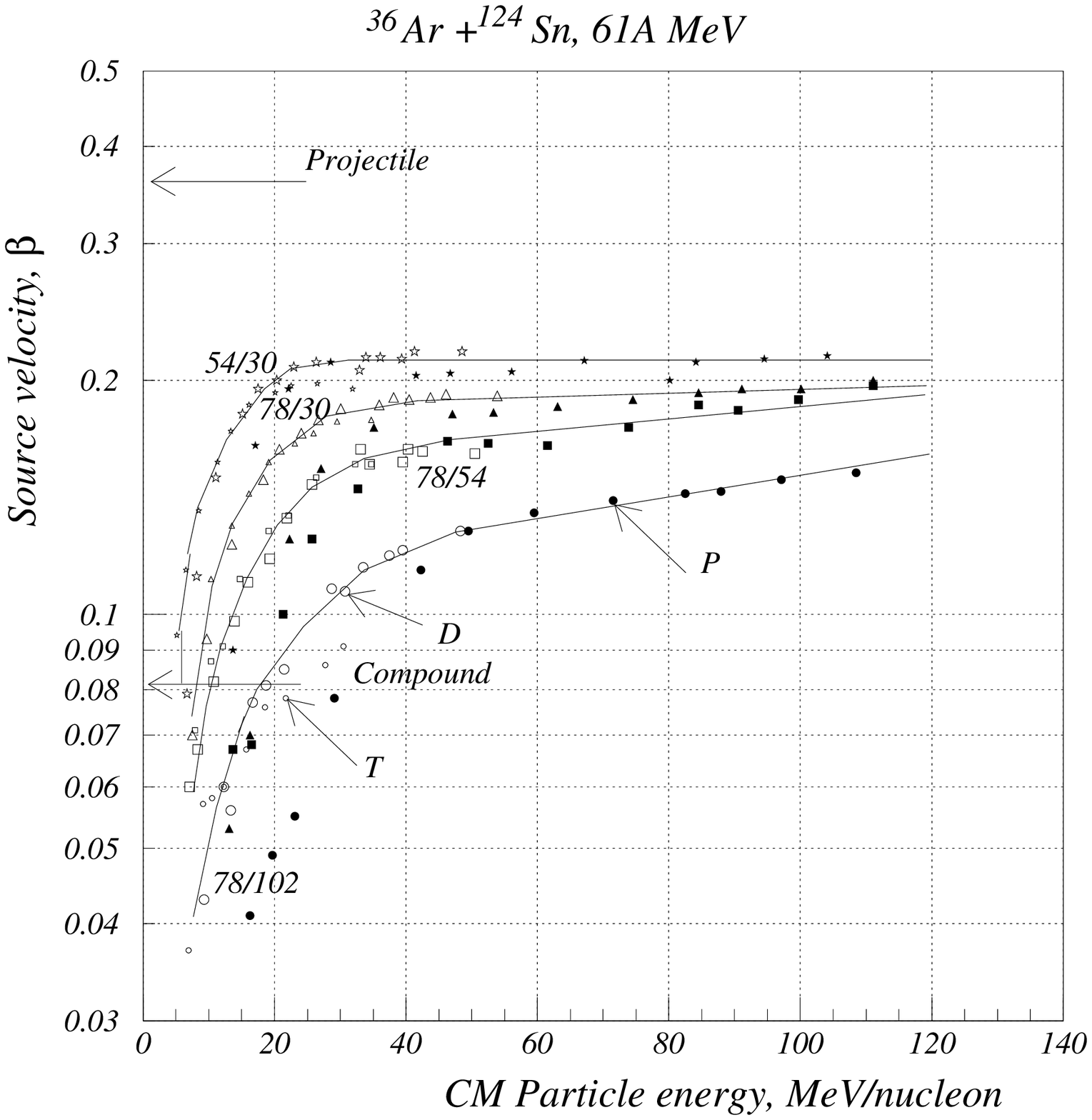,height=10cm,angle=0}}
\vspace{-2.cm}
\caption{\small
  The ``effective'' source velocity $\beta_{source}$ versus the kinetic energy 
  (in the source frame) of 
  the protons (full symbols), deuterons (big open symbols) and 
  tritons (small open symbols) 
  emitted in 61 MeV/nucleon $^{36}$Ar + $^{112}$Sn (upper panel) 
  and $^{36}$Ar + $^{124}$Sn (lower panel) collisions 
  for 4 pairs of angles. The solid lines are drawn to guide the eye.
}
\label{beta-tcm-sn}
\end{figure}

The middle panel assumes particle emission from two sources (QT and IS). 
It is useful to compare this plot with those of Fig.\ \ref{beta-tcm-sn}. 
One can see that, indeed, particle measured in the Ar + Sn reactions can 
be attributed to the contribution of two main sources, a QT and an IS.  

\begin{figure}
\vspace{3.cm}
\centerline{\psfig{file=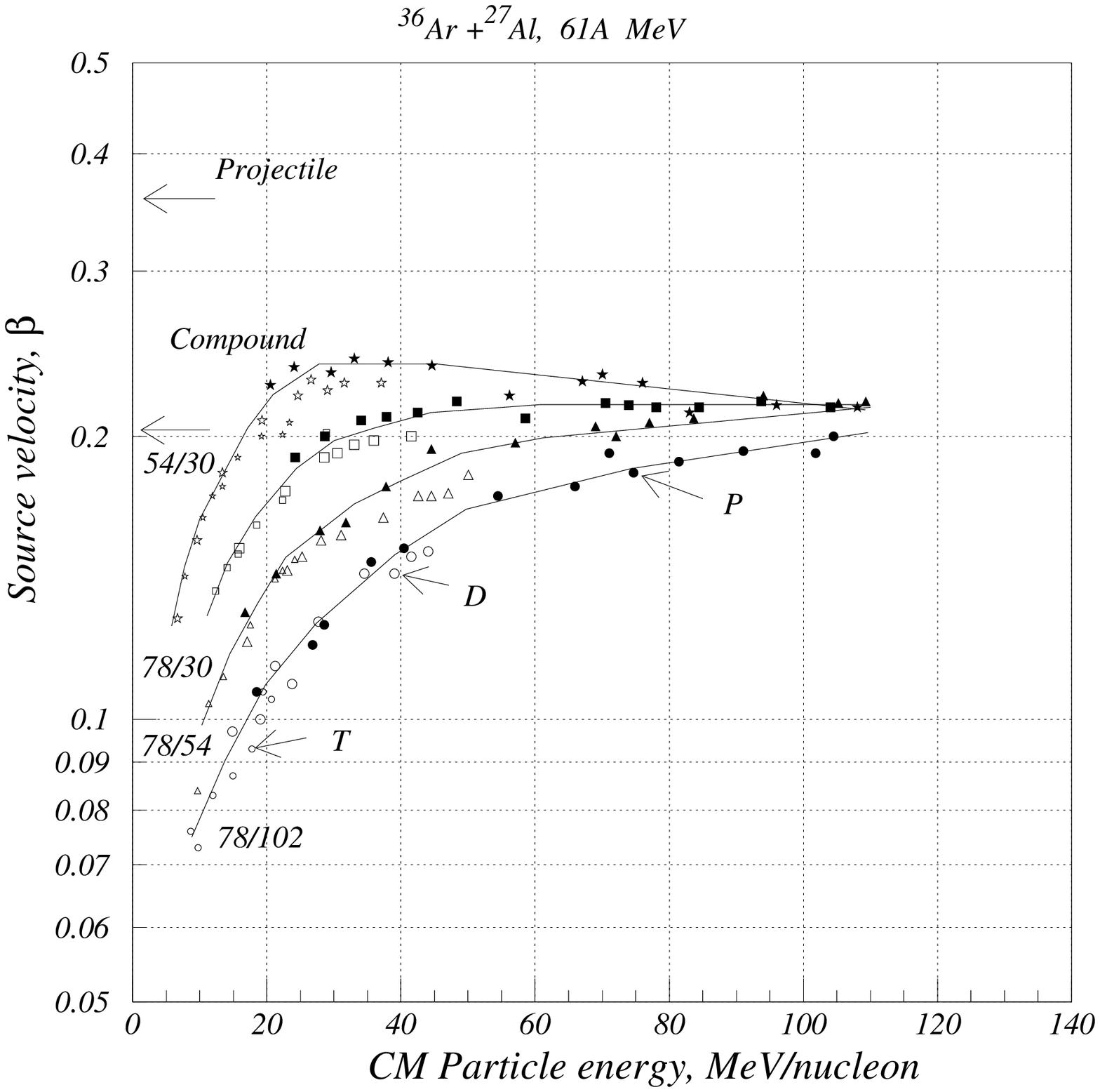,height=14.cm,angle=0}}
\vspace{-2.cm}
\caption{\small
  The ``effective'' source velocity  $\beta_{source}$ versus the kinetic energy 
  (in the source frame) of 
  the protons (full symbols), deuterons (big open symbols) and 
  tritons (small open symbols) 
  emitted in 61 MeV/nucleon $^{36}$Ar + $^{27}$Al collisions 
  for 4 pairs of angles. The solid lines are drawn to guide the eye.
}
\label{beta-tcm}
\end{figure}

Finally, the bottom panel assumes particle emission from all three 
sources (QT, QP and IS) as parameterized in \cite{Lanza98}. 
Comparing this plot with the Ar + Al data in Fig.\ \ref{beta-tcm} 
one can notice that:

1) The QP limiting region is much more defined in Fig.\ \ref{tano}
as compared to the experimental data (Fig.\ \ref{beta-tcm}). 
This is due to the inclusion of more forward angles in Fig.\ \ref{tano}.

2) The lines in Fig.\ \ref{tano} converge much earlier to a common value of
$\beta_{source}$, as $E_{CM}$ is increased, while the experimental data 
(Fig.\ \ref{beta-tcm}) present 
a much larger spreading in the intermediate $E_{CM}$ region. 
Thus, a broad range of impact parameters leads to a ``continuum'' 
of sources, while the standard three-moving sources parameterization 
performs an averaging over several sources.

\newpage

\newpage

\newpage

\newpage

\newpage

\end{document}